# Freestanding Resist Metasurface Supporting Higher-Order BICs for Efficient Field Enhancement in TMD Monolayers

*Chih-Zong Deng[1], Sunhao Shi[2], Chun-Hao Chiang[1], Mu-Hsin Chen[1], Jui-Han Fu[2], Vincent Tung[2], and Ya-Lun Ho\*[1]*

[1]Research Center for Electronic and Optical Materials, National Institute for Materials Science (NIMS), 1-1 Namiki, Tsukuba, Ibaraki 305-0044, Japan

[2]Department of Chemical System Engineering, School of Engineering, The University of Tokyo, 7-3-1 Hongo, Bunkyo, Tokyo 113-8656, Japan

**ABSTRACT.**

Enhancing light–matter coupling in two-dimensional (2D) semiconductors such as transition metal dichalcogenide monolayers remains a central challenge in nanophotonics due to their atomic thickness which limits their interaction volume with light. Here, we demonstrate that first-order quasi-bound states in the continuum (quasi-BICs) supported by a freestanding metasurface provide exceptionally strong surface field enhancement, enabling efficient coupling with a tungsten disulfide ($WS_2$) monolayer. Triangular-lattice polymer patterns on silicon nitride membranes are fabricated to realize these higher-order modes. Simulations reveal that first-order quasi-BICs exhibit much stronger field enhancement than zeroth-order modes at the top surface where the $WS_2$ monolayer is placed. Photoluminescence (PL) measurements confirm a remarkable PL enhancement factor of 127 for first-order quasi-BICs—over six times larger than that of zeroth-order quasi-BICs. These results establish higher-order BICs in freestanding metasurfaces as a powerful route to engineer light–matter interactions in 2D semiconductors for advanced nanophotonic and quantum photonic applications.

# 1. Introduction

Two-dimensional (2D) transition metal dichalcogenides (TMDs), such as $WS_2$ and $MoS_2$ monolayers, have emerged as a central platform for advancing next-generation optoelectronic and quantum technologies.[1,2] Their atomically thin nature provides unique opportunities for extreme miniaturization and integration, while their strong excitonic resonances and valley-selective optical selection rules open pathways toward exciton-based devices. These features have motivated intense research into TMD-based photodetectors, light-emitting devices, and quantum emitters, positioning them as key building blocks for future information and communication technologies.[3-8] However, a major bottleneck arises from the intrinsically weak interaction volume between light and monolayer semiconductors. Because TMD monolayers are atomically thin, their absorption and emission cross-sections are small compared to bulk semiconductors, severely limiting their efficiency when integrated into optical devices. Overcoming this limitation requires carefully engineered photonic environments that can amplify and control light–matter interactions at the nanoscale.

Nanophotonic resonators and metasurfaces offer an attractive solution to this challenge. By confining electromagnetic fields into subwavelength regions and enhancing their intensity, resonant photonic structures can compensate for the low optical volume of TMD monolayers. This strategy has already been demonstrated in various contexts, including photonic crystals,[4] plasmonic nanoantennas,[6] and dielectric metasurface.[9] Among these approaches, dielectric metasurfaces are particularly promising due to their compatibility with large-area active areas, low optical losses, and ability to sustain a rich set of resonant modes with tailored spectral and spatial characteristics.

In this context, bound states in the continuum (BICs) supported by dielectric metasurfaces have recently emerged as an especially powerful concept for enhancing light–matter interactions.[10, 11] BICs are non-radiating states of modes that remain perfectly confined within photonic systems, despite being embedded in the radiation continuum. When perturbed by symmetry breaking or fabrication imperfections, they transform into quasi-BICs with extremely high but finite quality (Q-) factors. These quasi-BIC metasurfaces produce strong field confinement and narrow resonances, enabling applications in lasing,[12-14] sensing,[15] nonlinear optics,[16] and quantum light generation.[17] The ability of quasi-BIC metasurfaces to concentrate light into subwavelength regions with minimal radiative loss makes them suited for coupling with atomically thin materials such as TMD monolayers.[18-25]

A critical design consideration in enhancing light–matter interaction is whether the quasi-BIC metasurface is supported on a substrate or suspended as a freestanding structure. Substrate-supported structures are easier to fabricate and integrate but suffer from several limitations: the lower refractive index contrast between the substrate and the nanostructure often leads to radiation leakage into the substrate, reducing the achievable Q-factors; furthermore, the substrate can perturb the field distribution, reducing the overlap of resonant modes with surface-bound emitters.[26, 27] In contrast, freestanding membranes eliminate substrate leakage pathways and allow for symmetric light confinement on both sides of the structure.[18, 28] This symmetry not only maximizes the achievable Q-factor but also promotes stronger near-field intensities. For atomically thin TMDs, which interact primarily through their immediate interface, freestanding platforms therefore provide a uniquely advantageous environment for enhancing excitonic emission.

Most prior studies have focused on exploiting fundamental (zeroth-order) BIC modes. These fundamental modes, which typically concentrate fields inside the photonic slab, have been used to enhance spontaneous emission, exciton–polariton formation, and nonlinear optics in TMDs.[18-25] However, reports on higher-order BICs remain limited.[13, 14] Because of their distinct field distributions, odd-order BIC modes exhibit stronger field localization near the surface region, thereby improving coupling efficiency with material placed on the surface. Such surface-localized resonances, which exhibit strong field enhancement on the top surface, are especially advantageous for TMD monolayers, enabling efficient coupling where the material is placed. Despite this compelling potential, systematic investigations into the role of higher-order BICs for enhancing TMD emission remain scarce.

Here, we experimentally demonstrate that first-order quasi-BICs supported by a freestanding metasurface—comprising a polymer hole-array slab on a silicon nitride (SiN) membrane—enable dramatically stronger photoluminescence (PL) enhancement in tungsten disulfide ($WS_2$) monolayers on top of the metasurface. Through field enhancement simulations and PL measurements, we identify the field distributions of first-order modes and confirm their decisive role in enhancing the field on the top surface. Our results demonstrate first-order BICs in freestanding metasurfaces as a powerful new design principle for boosting light–matter interaction in 2D materials. This approach can offer a general platform for engineering strong excitonic effects, nonlinear responses, and quantum optical functionalities, opening new directions for hybrid photonic–2D material systems.

## 2. Results and Discussion

Figure 1 illustrates the freestanding metasurface designed to strongly enhance the field near the top surface, thereby significantly enhancing the PL of a TMD monolayer at the surface. This enhancement is achieved by leveraging the first-order quasi-BICs. As shown in Figure 1a, the proposed device consists of a freestanding metasurface—a polymer hole-array slab on a SiN freestanding membrane. This specific geometry is critical for supporting the desired first-order quasi-BICs. The simulated electric energy density $E_{den}$ profiles of proposed freestanding metasurface as shown in Figure 1b and c visually confirm the strong field confinement. The first-order quasi-BIC in Figure 1b shows a strong field enhancement at the top surface of the metasurface, precisely where the monolayer is located. This strong overlap between the field and the exciton-rich region of the monolayer is essential for achieving efficient exciton-photon coupling. In contrast, the zeroth-order quasi-BIC in Figure 1c exhibits a weaker and more delocalized field at the top surface, leading to a less effective interaction. Figure 1d shows the PL spectrum of the $WS_2$ monolayer on the first-order quasi-BIC metasurface (red curve), which shows a remarkable PL enhancement—a dramatic increase in intensity compared to the PL from the same $WS_2$ monolayer on the zeroth-order quasi-BIC metasurface (light red curve) or the $WS_2$ monolayer on the unstructured membrane (yellow curve). The strongly enhanced PL confirms that the first-order quasi-BIC significantly boosts the radiative efficiency of the $WS_2$ monolayer, a direct consequence of the strongly enhanced field and efficient exciton-photon coupling.

Figure 2a shows a schematic diagram of the proposed freestanding metasurface, consisting of a triangular-lattice air-hole array patterned in polymer resist with a lattice period $P = 560$ nm, a hole diameter $D = 280$ nm, and a thickness $T = 400$ nm on a 50-nm-thick SiN freestanding membrane. This freestanding metasurface is designed to support guided modes (GMs) and BICs, with first-order modes in particular providing a strong field enhancement for the light-matter interactions with a $WS_2$ monolayer. The simulated angle-resolved reflectance spectra of the freestanding metasurface are shown in Figure 2b. The light is *x*-polarized illumination along the *y*-direction, calculated using rigorous coupled-wave analysis (RCWA) to characterize the structure. The incident angle $\theta_y$ is varied between 0° and 2.5° to resolve symmetry-protected BICs expected at the Γ point. At normal incidence, the spectrum exhibits vanishing linewidths for BIC modes, allowing clear distinction from GMs. In the longer wavelength range, four zeroth-order modes are identified, including one GM ($GM_{0th}$) and three BICs ($BIC_{0th,1}$, $BIC_{0th,3}$, $BIC_{0th,4}$). In the shorter wavelength range, four first-order modes are identified, including one GM ($GM_{1st}$) and three BICs ($BIC_{1st,1}$, $BIC_{1st,3}$, $BIC_{1st,4}$). In addition, $BIC_{0th,2}$ and $BIC_{1st,2}$ are accessible under *x*-tilted illumination ($\theta_x$), as shown in Figure S1. In practice, perfect BICs transform into quasi-BICs due to inevitable radiation leakage arising from fabrication imperfections, finite-size effects, and nonzero incident angles. Nevertheless, they maintain exceptionally high Q-factors, enabling strong light confinement. The spatial distributions of the $E_{den}$ for the zeroth-order and first-order modes at $\theta_y = 2.5°$ are shown in Figure 2c and d, respectively. The top panels show the in-plane (*xy*-plane) field profiles at the middle of the SiN membrane, while the bottom panels show the out-of-plane (*yz*-plane) cross-sections through the unit cell. The *xy*-plane field profiles maps confirm that the zeroth- and first-order modes share similar in-plane symmetries, consistent with their transverse electric nature (Figure S2). In contrast, the *yz*-plane field distributions highlight a distinct difference in vertical mode order: the zeroth-order mode exhibits a single antinode confined within the SiN membrane, whereas the first-order mode features two antinodes—one within the membrane and another near the resist–air interface. This strong field enhancement on the top surface highlights the potential of first-order BICs as an ideal environment for coupling with TMD monolayers, enabling efficient PL enhancement and exciton–photon interactions.

To highlight the importance of the freestanding architecture for accessing first-order modes with strong surface field confinement, Figure 3 compares the substrate-supported metasurface and the freestanding metasurface, further contrasting the field enhancement ~~characteristics~~ of zeroth- and first-order modes. Figures 3a and b show the simulated reflectance spectra, both at an incident angle $\theta_y = 2.5°$, under varying lattice period for polymer hole-array slabs on a silicon dioxide ($SiO_2$) substrate and the same slab on a SiN freestanding membrane, respectively. In both cases, the resonance wavelengths redshift with increasing lattice period, consistent with the enlarged effective optical path. Notably, the resonant wavelength can be readily tuned across the visible range by adjusting the lattice period, allowing precise spectral alignment with the excitonic transition of TMD monolayers to maximize light–matter interaction through exciton–photon coupling. As shown in Figure 3b, both zeroth- and first-order modes are clearly observed in the freestanding metasurface. In contrast, in the substrate-supported metasurface (Figure 3a), substrate-induced leakage strongly suppresses the formation of first-order GMs and quasi-BICs. Moreover, significant field leakage into the substrate leads to resonance attenuation, resulting in reduced reflectance contrast.

Figure 3c-e present the spatial distribution of the $E_{den}$ for the GM, $BIC_2$ and $BIC_4$ at $\theta_y = 2.5°$. The resonance wavelengths of the modes were tuned near the excitonic emission peak wavelength of $WS_2$ (~620 nm) by setting the lattice periods $P$ to 480 nm and 560 nm for the zeroth- and first-order modes, respectively. The top panels display the in-plane ($xy$-plane) field profiles at the top surface, while the bottom panels show the out-of-plane ($xz$-plane) cross-sections through the unit cell. For the substrate-supported case (left panel of Figure 3c-e), although the field is strongly confined within the SiN membrane (up to $10^4$ for $BIC_{0th,4}$), the enhancement at the top surface is reduced by approximately two orders of magnitude, resulting in weak field localized at the top surface. All modes in the substrate-supported case exhibit significantly weaker field enhancement compared to those in the freestanding metasurfaces (middle and right panel), revealing pronounced leakage into the substrate.

In contrast, the freestanding metasurfaces enable stronger field enhancement near the top surface. The zeroth-order modes show enhanced surface fields compared to the substrate-supported metasurface, while the first-order modes exhibit even stronger enhancement. Comparing the surface field enhancement among zeroth- and first-order modes, the maximum $E_{den}$ for $GM_{1st}$, $BIC_{1st,2}$, and $BIC_{1st,4}$, are 82, 444, and 3177, respectively—much larger than their zeroth-order counterparts $GM_{0th}$ (11), $BIC_{0th,2}$ (175), and $BIC_{0th,4}$ (1541). This enhancement arises because the antinodes of the field in first-order modes coincides with the near top surface (bottom panel). Furthermore, when comparing quasi-BICs with GMs, the quasi-BICs consistently show stronger enhancement, with $BIC_{1st,4}$ reaching more than an order of magnitude higher than $GM_{1st}$ (right panel of Figures 3c and e). In summary, substrate-induced leakage suppresses the formation of first-order modes, whereas the freestanding configuration preserves them. Owing to their field distributions, the first-order modes provide substantially stronger surface field enhancements than the zeroth-order modes, directly benefiting light–matter interactions with TMD monolayers.

The designed freestanding metasurface was fabricated for experimental characterization of its optical properties (optical microscope (OM) and scanning electron microscope images of fabricated freestanding metasurfaces as shown in Figures S3 and S4, respectively). Since the SiN membrane used in this work exhibits defect-related states, a broad-band PL emission is observed under optical pumping (Figure S5), allowing the investigated GMs and quasi-BICs to be coupled to this PL band. Figure 4a presents experimental PL spectra from freestanding metasurfaces with varying lattice periods $P$ (420–580 nm) before the $WS_2$ monolayer was transferred. PL spectra reveal multiple emission peaks corresponding to optical resonances. As expected from simulations, the resonance wavelengths redshift with increasing lattice period due to the enlarged optical path. For the lattice period of 420 nm, a group of resonances appears near a wavelength of 550 nm, corresponding to zeroth-order modes, and shift to a wavelength of 700 nm for a 580-nm lattice period. At 460-nm and 480-nm periods, the zeroth-order modes occur around 620 nm, overlapping with the exciton emission wavelength of $WS_2$. In addition, first-order modes emerge at shorter wavelengths (~510 nm at 440 nm lattice period). For a 560-nm lattice period, they overlap with the exciton emission wavelength of $WS_2$. For GMs, the $GM_{0th}$ at 590.2 nm with a Q-factor of 138 (460 nm lattice period) and $GM_{1st}$ at 582.6 nm with a Q-factor of 55 (560 nm lattice period) near the exciton emission wavelength $WS_2$. The PL spectra with a narrow range to highlight the zeroth- and first-order quasi-BICs are shown in Figures 4b and 4c. For a 460-nm period (Figure 4b), the multi-peak group is decomposed into Lorentzian components corresponding to $BIC_{0th,1}$, $BIC_{0th,2}$, and $BIC_{0th,3}$ at 598.7, 601.5, and 605.4 nm with Q-factors of 214, 143, and 162, respectively. For a 560-nm lattice period (Figure 4c), the first-order group decomposes into $BIC_{1st,1}$, $BIC_{1st,2}$, and $BIC_{1st,3}$ at 607.7, 613.7, and 616.9 nm with Q-factors of 122, 173, and 251, respectively. Simulated reflectance spectra for the corresponding periods, as shown in Figure 4d and e, reproduce the experimental trends. For the 460-nm lattice period (Figure 4d), reflectance peaks occur at 572.3, 576.3, and 579.4 nm, while for the 560-nm lattice period (Figure 4e), peaks appear at 592.1, 598.6, and 602.2 nm. The shifts between experiment and simulation are attributed to deviations in the refractive indices of SiN and the polymer resist. These results experimentally demonstrate that the freestanding metasurface architecture can supports both zeroth- and first-order quasi-BICs across the visible spectrum by tunning the lattice period, providing a versatile platform for enhancing light–matter interactions with 2D materials such as TMD monolayers.

To verify the PL enhancement arising from field enhancement by the optical modes, $WS_2$ monolayers were transferred onto the freestanding metasurfaces (see Figure S3 for OM images). Figure 5a and b present the PL spectra obtained from freestanding metasurface supporting zeroth-and first-order modes, respectively. In Figure 5a, freestanding metasurfaces with lattice periods $P$ = 440, 460, and 480 nm supporting zeroth-order modes that spectrally overlap with the $WS_2$ exciton emission band. The emission from $WS_2$ on an unstructured membrane is shown for reference (scaled ×10 for visibility). Among these freestanding metasurfaces, the one with $P$ = 460 nm presents the strongest enhancement because its quasi-BICs overlap most closely with the intrinsic PL emission peak of $WS_2$. For $P$ = 440 nm, the resonance is spectrally detuned, resulting in weaker enhancement. At $P$ = 480 nm, although the $GM_{0th}$ overlaps spectrally with the $WS_2$ PL emission peak, its field confinement is weaker than that of quasi-BICs, leading to relatively modest enhancement compared to the $P$ = 460 nm case. Figure 5b shows metasurfaces with $P$ = 540, 560, and 580 nm, supporting first-order modes that spectrally overlap with the $WS_2$ emission band. For $P$ = 540 nm, the resonance is spectrally detuned from the $WS_2$ emission band, leading to a weaker enhancement. Nevertheless, the PL intensity remains comparable to that of the $P$ = 460 nm freestanding metasurface, where the emission is enhanced by zeroth-order modes. The most pronounced enhancement occurs for $P$ = 560 nm, where the first-order quasi-BICs are spectrally aligned with the $WS_2$ emission peak. Thanks to the precise spectral overlap between the first-order quasi-BICs and the $WS_2$ emission band achieved by tuning the lattice period, as well as the high spatial overlap between the enhanced field and the $WS_2$ monolayer arising from the distinct field distribution of the first-order mode, efficient exciton–photon coupling is facilitated, resulting in a markedly intensified PL peak. At $P$ = 580 nm, enhancement remains significant even though it originates primarily from $GM_{1st}$ with weaker field enhancement rather than from quasi-BICs, underscoring the stronger surface field confinement of first-order modes compared to zeroth-order modes. Notably, the three peaks near 630 nm correspond to first-order quasi-BICs. Despite their detuning from the $WS_2$ PL maximum, they still drive substantial emission enhancement, further evidencing the robust surface coupling provided by first-order resonances. Figure 5c directly compares PL enhancement factors for freestanding metasurface with $P$ = 460 nm (PL enhanced by zeroth-order modes) and $P$ = 560 nm (PL enhanced by first-order modes). The enhancement factor, defined as the PL intensity ratio between metasurfaces and the unstructured membrane, reaches 127 for the first-order quasi-BICs at $P$ = 560 nm. Even the weaker $GM_{1st}$ resonance at ~585 nm yields an enhancement factor of 90, far exceeding those of zeroth-order modes. In contrast, the peak enhancement factor for $P$ = 460 nm (zeroth-order mode) is only 21. This difference confirms that the stronger field enhancement on the top surface of first-order resonances directly leads to more efficient exciton–photon coupling in $WS_2$ monolayer.

## 3. Conclusions

We have demonstrated that freestanding metasurfaces can host both zeroth- and first-order quasi-BICs across the visible spectrum, enabling efficient coupling with $WS_2$ monolayers. While zeroth-order BICs primarily confine fields inside the freestanding metasurface, first-order BICs provide strong field enhancement on the top surface of the freestanding metasurface that overlaps with the $WS_2$ monolayer. As a result, first-order resonances provide significantly stronger PL enhancement—exceeding six-fold higher than zeroth-order counterparts. Our experimental and simulation results demonstrate first-order BICs in freestanding metasurfaces as a powerful design principle for enhancing light–matter interactions in 2D materials such as TMD monolayer. Beyond PL enhancement, this approach opens new pathways for engineering strong exciton–photon coupling, nonlinear processes, and quantum optical functionalities in 2D material–based nanophotonic devices.

## 4. Methods
**Numerical Simulations**

The far-field reflectance spectra of the freestanding metasurfaces and the near-field electric field profiles their resonant modes were numerically analyzed using the rigorous coupled-wave analysis method (DiffractMOD, RSoft Design Group, USA). The simulations employed periodic boundary conditions along the the *x*- and *y*-directions, and perfectly matched-layer boundaries along the *z*-direction. A plane wave was used as the incident source, propagating along the z-axis. The electric field **E** was normalized to the amplitude of the incident field. The electric energy density was calculated as $U_E = \frac{1}{2}\int \text{Re}[\varepsilon(\mathbf{r}')]|\mathbf{E}|^2 \, dV$, where **E** is the electric field, $\varepsilon$ is the spatially dependent permittivity, and V is the volume of the simulation domain.

**Fabrication**

A triangular lattice hole-array pattern was defined in the polymer resist on the SiN membrane using electron-beam lithography. The SiN membranes were provided by Ted Pella, Inc.

**Preparation for WS$_2$ monolayer**

WS$_2$ films were grown via standard chemical vapor deposition (CVD) using a one-zone horizontal quartz tube furnace (2-inch diameter). Tungsten trioxide (WO$_3$, Sigma-Aldrich, 99.9%, 100 mg) served as the tungsten source and was placed in the central heating zone along with c-plane sapphire substrates. The central zone was heated to 960 °C and maintained for 15 min to enable film growth. Sulfur powder (S, Sigma-Aldrich, 99.99%, 3 g), pre-solidified prior to use, was positioned upstream in a quartz boat and heated to 145 °C using an external heating belt. Sulfur heating began 10 min before the center zone reached the target temperature. Prior to the growth process, the quartz tube was evacuated to base pressure and purged with a carrier gas mixture of Ar (200 sccm) and H$_3$ (40 sccm) at a pressure of 20 torr. After growth, the furnace was naturally cooled to room temperature while maintaining the Ar/H$_3$ flow.

Transfer of $WS_3$ films was achieved through a polydimethylsiloxane (PDMS)-assisted wet-etching technique. A home-prepared PDMS stamp was laminated onto the $WS_2$ surface by natural adhesion. The sapphire substrate was subsequently etched away in a KOH solution, releasing the $WS_2$ film onto the PDMS. The PDMS/$WS_2$ assembly was then aligned and brought into contact with the freestanding metasurface, followed by vacuum treatment for 1 h to improve adhesion. Finally, gentle heating at 60 °C for 10 min facilitated detachment of the PDMS, leaving the $WS_2$ film successfully transferred onto the metasurface.

**Optical Characterization**

The excitation light was focused onto the sample using a 10× objective lens. Photoluminescence (PL) spectra of the freestanding metasurfaces were measured with a confocal laser microscope system (alpha300 R, WITec, Germany). A continuous-wave (CW) laser with a wavelength of 488 nm served as the excitation source. The beam was focused onto the sample through a 10× objective lens (NA = 0.25), and the emitted PL signals were collected through the same objective and analyzed using a spectrometer.

**FIGURES.**

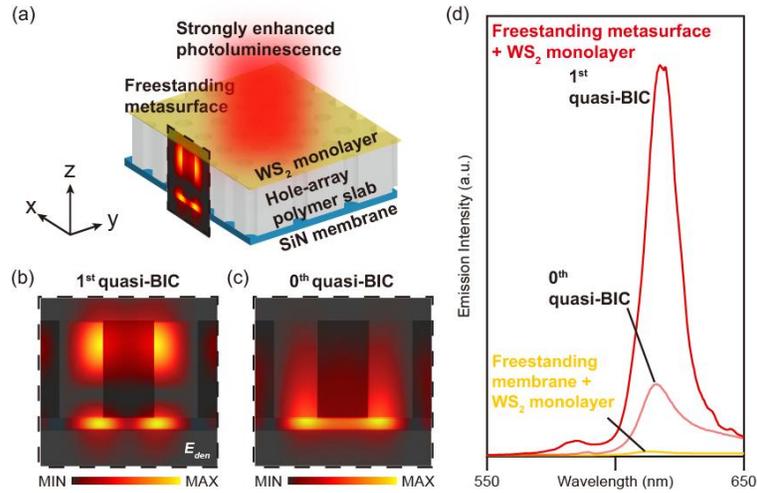

**Figure 1.** Photoluminescence (PL) from a tungsten disulfide (WS$_2$) monolayer strongly enhanced by a freestanding metasurface supporting first-order quasi-bound states in the continuum (quasi-BICs), which provide intense surface field confinement and efficient exciton–photon coupling. (a) Schematic illustration of the concept: A WS$_2$ monolayer is placed on a freestanding metasurface, which is a hole-array polymer slab on a silicon nitride (SiN) membrane. Simulated electric energy density $E_{den}$ distribution at (b) first-order quasi-BIC and (c) zeroth-order quasi-BIC in *yz*-plane. (d) PL spectra of the WS$_2$ monolayer under different conditions: first-order quasi-BIC (red curve), zeroth-order quasi-BIC (light red curve), and on an unstructured membrane (yellow curve).

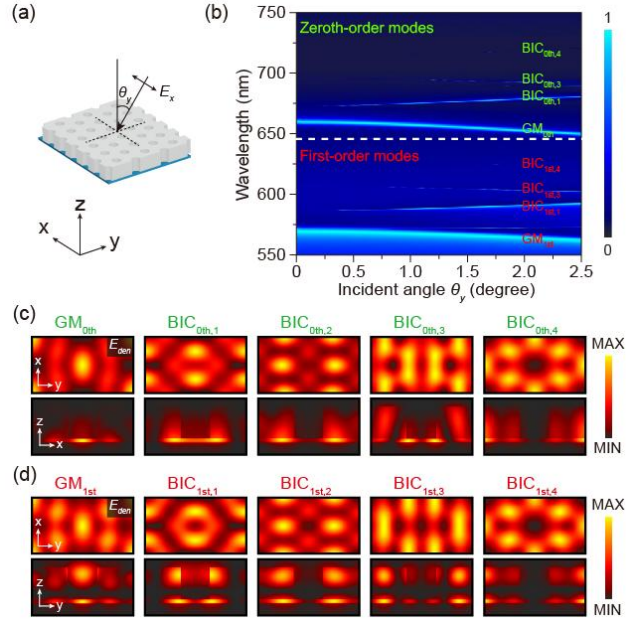

**Figure 2.** Freestanding metasurface engineered to support first-order guided modes (GMs) and quasi-BICs with strong field enhancement at the top surface. (a) Schematic diagram of the membrane metasurface, consisting of a triangular-lattice air-hole array patterned in polymer resist with a lattice period $P = 560$ nm, a hole diameter $D = 280$ nm, and a thickness $T = 400$ nm on a 50-nm-thick SiN membrane. (b) Simulated angle-resolved reflectance spectra under $x$-polarized illumination along the $y$-direction, showing including GMs and quasi-BICs. Electric energy density $E_{den}$ distributions for the (c) zeroth-order and (d) first-order modes, the incident angle $\theta = 2.5°$.

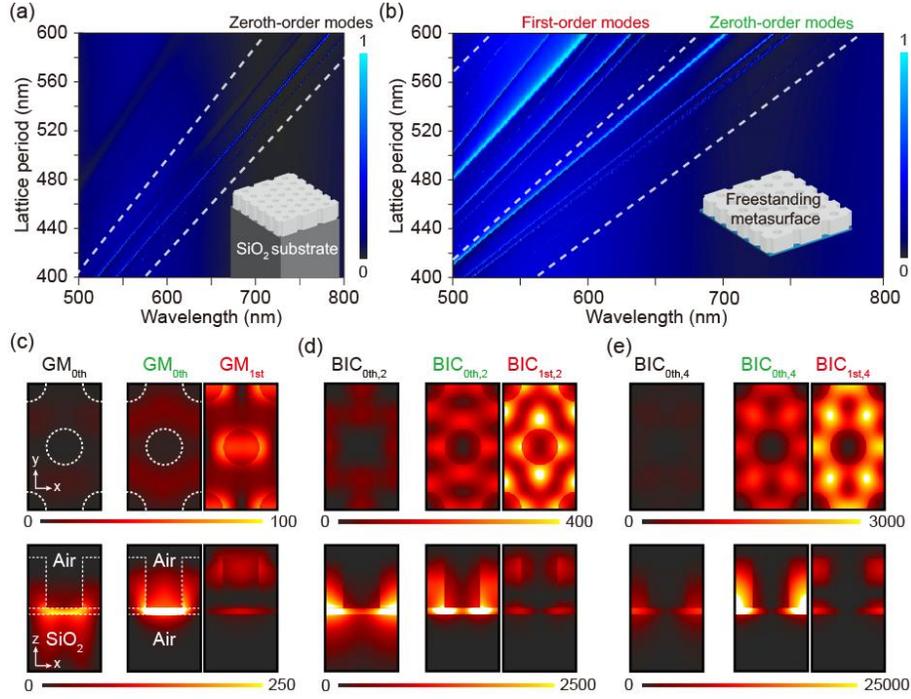

**Figure 3.** Strong field enhancement on the top surface of the freestanding metasurface via first-order modes. Simulated reflectance spectra as a function of lattice period for (a) substrate-based metasurface and (b) freestanding metasurface. Electric energy density $E_{den}$ distributions in the *xy*-plane at the top surface and *xz*-plane for (c) GM, (d) $BIC_2$, and (e) $BIC_4$ in both the substrate-based metasurface and freestanding metasurface. The incident light is *x*-polarized with an incident angle of 2.5° along the *y*-direction.

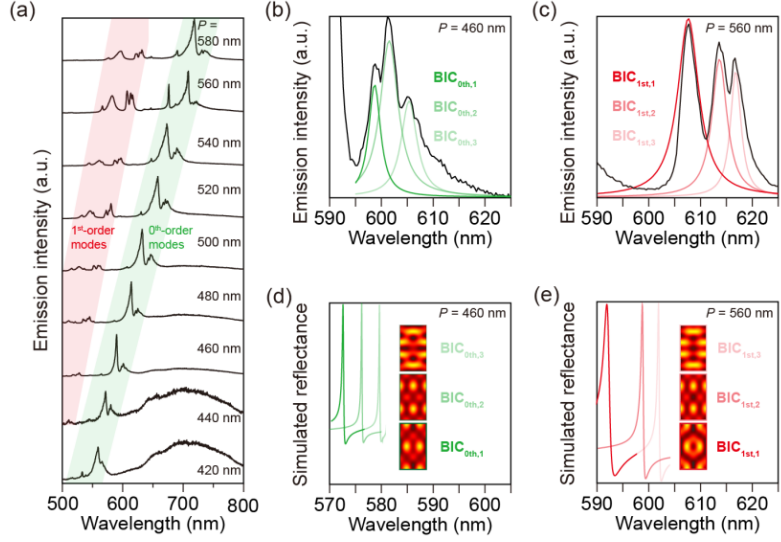

**Figure 4.** Demonstration of both zeroth-order and first-order GMs and quasi-BICs by freestanding metasurfaces. (a) PL spectra of the freestanding metasurfaces as a function of lattice period. (b, c) PL spectra highlighting the spectral regions corresponding to the zeroth-order ($P$ = 460 nm) and first-order ($P$ = 560 nm) quasi-BICs, respectively. (d, c) Simulated reflectance spectra of freestanding metasurfaces with $P$ = 460 nm and $P$ = 560 nm, respectively. Insets show the $E_{den}$ distributions of the quasi-BICs.

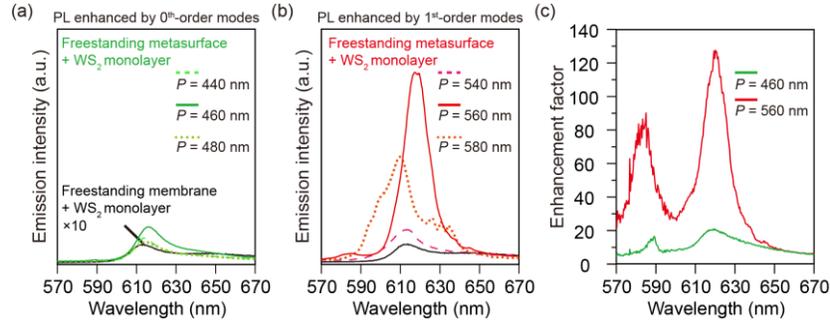

**Figure 5.** Enhanced PL emission from WS$_2$ monolayers via GMs and quasi-BICs in the freestanding metasurfaces. (a, b) PL spectra of WS$_2$ monolayers on freestanding metasurfaces supporting zeroth- and first-order modes, with lattice periods of 440, 460, and 480 nm (zeroth-order modes) and 540, 560, and 580 nm (first-order modes). The PL spectrum from WS$_2$ on the unstructured freestanding membrane is shown for reference. Both zeroth- and first-order modes are coupled to the WS$_2$ excitonic emission. (c) Enhancement factor, defined as the PL intensity ratio between the WS$_2$ monolayer on the freestanding metasurfaces and that on the unstructured membrane, for freestanding metasurfaces with lattice periods of 460 and 560 nm.

## AUTHOR INFORMATION

**Corresponding Author**

*E-mail: Ya-Lun Ho, HO.Ya-Lun@nims.go.jp

**Author Contributions**

The manuscript was written through contributions of all authors. All authors have given approval to the final version of the manuscript.

**Funding Sources**

JSPS KAKENHI Grant Number JP23K26155, JP25KF0083, Advanced Research Infrastructure for Materials and Nanotechnology in Japan (ARIM) of the Ministry of Education, Culture, Sports, Science and Technology (MEXT) (Proposal Number 25NM5090)

## ACKNOWLEDGMENT

This work was supported by JSPS KAKENHI Grant Number JP23K26155, JP25KF0083. A part of this work was supported by Advanced Research Infrastructure for Materials and Nanotechnology in Japan (ARIM) of the Ministry of Education, Culture, Sports, Science and Technology (MEXT) (Proposal Number JPMXP1225NM5090).